\documentclass{ws-ijmpe}

\begin{document}

\markboth{S. Ahmad, M. Bhuyan and S. K. Patra}{Properties of Z=120 and decay half-life..}

\catchline{}{}{}{}{}

\title{Properties of Z=120 nuclei and the $\alpha$-decay chains 
of the $^{292,304}$120 isotopes using relativistic and non-relativistic 
formalisms\\
}

\author{Shakeb Ahmad}
\address{Department of Physics, Aligarh Muslim University, Aligarh-
202 002, India.
\footnote{S. Ahmed,
Email: physics.sh@gmail.com}\\
}

\author{M. Bhuyan}
\address{School of Physics, Sambalpur University, Jyotivihar, Burla-768019, 
India.
\footnote{M. Bhuyan,
Email: bunuphy@iopb.res.in}\\
}

\author{S. K. Patra}
\address{Institute of Physics, Sachivalaya Marg,
Bhubaneswar-751 005, India.
\footnote{Dr. S. K. Patra,
Email: patra@iopb.res.in}\\
}

\maketitle

\begin{history}
\received{(received date)}
\revised{(revised date)}
\end{history}

\begin{abstract}
The ground state and first intrinsic excited state of superheavy
nuclei with Z=120 and N=160-204 are investigated using both
non-relativistic Skyrme-Hartree-Fock and the axially deformed
Relativistic Mean Field formalisms. We employ a simple BCS
pairing approach for calculating the energy contribution from 
pairing interaction. The results for isotopic chain of binding 
energy, quadrupole deformation parameter, two neutron separation 
energies and some other observables are compared with the FRDM 
and some recent macroscopic-microscopic calculations. We predict 
superdeformed ground state solutions for almost all the isotopes. 
Considering the possibility of magic neutron number, two different 
mode of $\alpha$-decay chains $^{292}$120 and $^{304}$120 are also 
studied within these frameworks. The $Q_\alpha$-values and the 
half-life $T_{1/2}^{\alpha}$ for these two different mode of
decay chains are compared with FRDM and recent macroscopic-microscopic
calculations. The calculation is extended for the $\alpha$-decay chains
of $^{292}$120 and $^{304}$120 from their exited state configuration
to respective configuration, which predicts long half-life
$T_{1/2}^{\alpha}$(sec.).
\end{abstract}

\section{Introduction}

By superheavy elements (SHEs) we mean elements with proton
number (Z) near the next magic number beyond the magic number
Z = 82, corresponding to Pb. The possibility of finding the 
magic or doubly magic isotopes of SHEs led to the prediction 
of a region of enhanced stability in the 1960's 
\cite{myers65,sobi66,meldner67,meldner67a,mosel69}. 
Since then,the nuclear synthesis and investigation of new 
superheavy elements has been a challenging problem in nuclear 
physics. A worldwide effort has been made to explore the island
of stability of SHEs. Many experimental groups in various 
laboratories are trying hard to study various peculiar aspects 
of SHEs 
\cite{ikez98,hof00,gen04,sto07,herz08,herz08a,herz08b,hof09,mor09,stav09,ket09,oga11}. 
For the synthesis of heavy and superheavy elements, two approaches 
have been successfully employed. Firstly, via cold fusion reactions, 
which have been successfully used to synthesize superheavy elements 
up to Z = 112 at GSI \cite{hof00,hof95,hoff95,hof96,hof98} and that 
with Z = 113 at RIKEN \cite{mor04}, and to confirm these experiments 
at RIKEN \cite{mor04,mor041,mor07} and LBNL \cite{gin03}. Secondly, 
hot fusion reactions have also been used to synthesize superheavy 
elements from Z = 112 to 116 and 118 \cite{og98,og01,og04,og07,og10,eic07}. 
Efforts are on to synthesize still heavier elements in various 
laboratories all over the world.

An impressive progress in the synthesis and experimental studies 
of the heaviest nuclei 
\cite{ikez98,hof00,gen04,sto07,herz08,herz08a,herz08b,hof09,mor09,stav09,ket09,oga11} 
requires intensive theoretical studies of them. Their studies 
are needed for predicting stability properties of as yet 
undiscovered SHEs and also for the interpretation of already 
existing experimental results. In the last decade, several 
theoretical investigations of SHEs are focused both on the 
structure and decay properties and on the synthesis mechanism
\cite{patyk91,nix94,nix94a,nilsson69,rutz97,rutz99,gupta97,patra1,sil04,cwiok96,cwiok96a,kruppa00,marinov07,marinov09,marinov09a,bhu09,lal96,long02,ren001,ren002,ren03,ren04,zha05,zha06,li03,zha061,fen07,she08,li09,zag10}.

In general most of the advanced model calculations predict
the existence of a closed shell at N = 184; however, they 
differ in predicting the atomic number of the closed proton 
shell. Some macroscopic-microscopic (MM) theories which 
traditionally involve a priory the knowledge of densities, 
single particle potentials and other bulk properties, they 
predict the magic shells at Z=114 and N=184 
\cite{patyk91,nix94,nix94a,mosel69,nilsson69}.
At the same time, the predictions of shell closure for the
superheavy region within the relativistic and non-relativistic
theories depend mostly on the force parameters 
\cite{rutz97,rutz99}. For example, the Skyrme Hartree-Fock 
(SHF) calculation with SkI4 force gives Z=114, N=184 as the 
next shell-closures and the relativistic microscopic mean
 field formalism (RMF) \cite{gupta97} predicts the probable 
shell-closures at Z=120 and N=184. Recently, more microscopic 
theoretical calculations have predicted the various other 
region of stability, beyond Z=82, N=126, as Z=120, N=172 or 
184 \cite{rutz97,gupta97,patra1} and Z=124 or 126, N=184 
\cite{cwiok96,cwiok96a,kruppa00}. Such estimations of structure properties
of nuclei in the superheavy mass region is a challenging area in
nuclear physics and a fruitful path towards the understanding
of 'island of stability' beyond the spherical doubly-magic
nucleus~\cite{cwi05,sob07}. Progress in understanding the structure
of the heaviest nuclei can be achieved through the theoretical
and experimental studies of production and decay of superheavy
elements (SHEs).

SHEs are an excellent testing ground for nuclear theory models. 
The SHE in the ground state is formed at the end of the 
cooling-down process of the compound nucleus. The $\alpha$ particles
are mainly emitted from the ground state of a formed SHE, because as
a rule, the $\gamma$-decay half-lives of low-lying excitation
states are shorter than the $\alpha$-decay half-lives of corresponding 
levels. Alpha decay is one of two main decay modes of the heaviest 
nuclei. It is important for these nuclei because: many of the already
known heavy nuclei decay by this mode, also many of the nuclei
not yet observed, specially superheavy nuclei are predicted to
be $\alpha$ emitters, and properties of this decay give a good
method for identification of decaying nuclei (genetic chain).
The amount of already collected data for $\alpha$-decay is quite 
large \cite{hof00,oga11,hof09,mor09,sto07,gen04,ikez98,stav09,ket09,herz08,herz08a,herz08b,hof95,hoff95,hof96,hof98,mor04,mor041,mor07,gin03,og98,og01,og04,og07,og10,eic07,tur03,audi03,wap03} 
and is still increasing. It is important then, both the
interpretation of existing data and for predictions for new
experiments, to realize with what accuracy one can presently
describe both observables of this process: $\alpha$-decay energy
$Q_\alpha$ and $\alpha$-decay half life $T_\alpha$. The $Q_\alpha$
energy is obtained  from masses of the respective nuclei,
which are presently described by a number of various methods 
\cite{patyk91,nix94,nilsson69,rutz97,gupta97,patra1,cwiok96a,kruppa00,marinov09,marinov09a,bhu09,lal96,long02,ren002,ren03,zha05,zha06,li03,she08,li09,zag10,moll97,moll97a,mye96,liran00,ton00,sobi89,mu01,muni01,muni03,muni004,ski10}.
Half-lives $T_{1/2}^\alpha$ are usually described 
in a phenomenological way. The very possibility of an extremely 
heavy {\it Z} nucleus motivated us to see the structure of such 
nuclei in an isotopic mass chain. Therefore, on the basis of the 
RMF and nonrelativistic SHF methods, we calculated the bulk 
properties of a Z = 120 nucleus in an isotopic chain of mass 
{\it A} = 280-324. This choice of mass range covers both the 
predicted neutron magic numbers {\it N} = 172 and 184.

The paper is organized as follows. Section II gives a brief 
description of the relativistic and nonrelativistic mean-field 
formalism. The pairing effects for open shell nuclei, included 
in our calculations, are also discussed in this section. The 
results of our calculation are presented in Section III, and 
Section IV includes the $\alpha$-decay modes of $^{292}$120 and
$^{3044}$120 isotopes. A summary of our results, together with 
the concluding remarks, are given in the last Section V.

\section{Theoretical Framework}
\subsection{The Skyrme Hartree-Fock Method}

The general form of the Skyrme effective interaction, used
in the mean-filed models, can be expressed as an energy density
functional $\mathcal{H}$ \cite{cha97,cha98,stone07},
\begin{eqnarray}
\mathcal{H}=\mathcal{K}+\mathcal{H}_0+\mathcal{H}_3+\mathcal{H}_{eff}+.... ,
\end{eqnarray}
where $\mathcal{K}=\frac{\hbar^2}{2m}\tau$ is the kinetic energy term with
$m$ as the nucleon mass, $\mathcal{H}_0$ is the zero range, $\mathcal{H}_3$
the density dependent term, and $\mathcal{H}_{eff}$ the effective-mass
dependent term, relevant for calculating the properties of
nuclear matter, are functions of nine parameters, $t_i$,
$x_i$ (i = 0, 1, 2, 3), and $\eta$, given as
\begin{eqnarray}
\mathcal{H}_0&=&\frac{1}{4}t_0\left[\left(2+x_0\right)\rho^2
-\left(2x_0+1\right)\left(\rho_p^2+\rho_n^2\right)\right]\\
\mathcal{H}_3&=&\frac{1}{24}t_3\rho^\eta\left[\left(2+x_3\right)\rho^2
-\left(2x_3+1\right)\left(\rho_p^2+\rho_n^2\right)\right]\\
\mathcal{H}_{eff}&=&\frac{1}{8}\left[t_1\left(2+x_1\right)
+t_2\left(2+x_2\right)\right]\tau\rho \nonumber\\
&+&\frac{1}{8}
\left[t_2\left(2x_2+1\right)
-t_1\left(2x_1+1\right)\right]\left(\tau_p\rho_n+\tau_n\rho_p\right).
\end{eqnarray}
The other terms, representing the surface contributions of a
finite nucleus with $b_4$ and $b_4^\prime$ as additional parameters, are
\begin{eqnarray}
\mathcal{H}_{S\rho}&=&\frac{1}{16}\left[3t_1\left(1+\frac{1}{2}x_1\right)
-t_2\left(1+\frac{1}{2}x_2\right)\right]\left(\vec{\nabla}\rho\right)^2 \nonumber\\
&&-\frac{1}{16}\left[3t_1\left(x_1+\frac{1}{2}\right)
+t_2\left(x_2+\frac{1}{2}\right)\right]\nonumber\\
&&\times\left[\left(\vec{\nabla}\rho_n\right)^2+
\left(\vec{\nabla}\rho_p\right)^2\right],
\end{eqnarray}
and
\begin{equation}
\mathcal{H}_{S\vec{J}}=-\frac{1}{2}\left[b_4\rho\vec{\nabla}\cdot\vec{J}
+b_4^\prime\left(\rho_n\vec{\nabla}\cdot\vec{J_n}+
\rho_p\vec{\nabla}\cdot\vec{J_p}\right)\right],
\end{equation}
here, the total nucleon number density $\rho=\rho_n+\rho_p$,
the kinetic energy density $\tau=\tau_n+\tau_p$, and the
spin-orbit density $\vec{J}=\vec{J_n}+\vec{J_p}$, with $n$ and $p$
referring to neutron and proton, respectively. The $\vec{J_q}=0$,
$q=n$ or $p$, for spin-saturated nuclei, i.e., for nuclei with
major oscillator shells completely filled. The total binding energy
(BE) of a nucleus is the integral of the energy density functional
$\mathcal{H}$. We have used here the Skyrme SkI4 and SLy4 sets
with $b_4\neq b_4^\prime$~\cite{rein95},
designed for considerations of proper spin-orbit interaction in finite
nuclei, related to the isotopic shifts in the Pb region.

\subsection{The Relativistic Mean-Field Formalism}
The relativistic Lagrangian density for a nucleon-meson many-body 
system \cite{sero86,ring90,ring90a,boguta},
\begin{eqnarray}
{\cal L}&=&\overline{\psi_{i}}\{i\gamma^{\mu}
\partial_{\mu}-M\}\psi_{i}
+{\frac12}\partial^{\mu}\sigma\partial_{\mu}\sigma
-{\frac12}m_{\sigma}^{2}\sigma^{2}\nonumber\\
&&-{\frac13}g_{2}\sigma^{3} -{\frac14}g_{3}\sigma^{4}
-g_{s}\overline{\psi_{i}}\psi_{i}\sigma-{\frac14}\Omega^{\mu\nu}
\Omega_{\mu\nu}\nonumber\\
&&+{\frac12}m_{w}^{2}V^{\mu}V_{\mu}
+{\frac14}c_{3}(V_{\mu}V^{\mu})^{2} 
-g_{w}\overline\psi_{i}\gamma^{\mu}\psi_{i}
V_{\mu}\nonumber\\
&&-{\frac14}\vec{B}^{\mu\nu}.\vec{B}_{\mu\nu}+{\frac12}m_{\rho}^{2}{\vec
R^{\mu}} .{\vec{R}_{\mu}}
-g_{\rho}\overline\psi_{i}\gamma^{\mu}\vec{\tau}\psi_{i}.\vec
{R^{\mu}}\nonumber\\
&&-{\frac14}F^{\mu\nu}F_{\mu\nu}-e\overline\psi_{i}
\gamma^{\mu}\frac{\left(1-\tau_{3i}\right)}{2}\psi_{i}A_{\mu} .
\end{eqnarray}
All the quantities have their usual well-known meanings. From the above
Lagrangian we obtain the field equations for the nucleons and mesons.
These equations are solved by expanding the upper and lower components
of the Dirac spinors and the boson fields in an axially deformed harmonic
oscillator basis, with an initial deformation $\beta_{0}$. The set of 
coupled equations is solved numerically by a self-consistent iteration 
method. The center-of-mass motion energy correction is estimated by the 
usual harmonic oscillator formula $E_{c.m.}=\frac{3}{4}(41A^{-1/3})$.
The quadrupole deformation parameter $\beta_2$ is evaluated from the 
resulting proton and neutron quadrupole moments, as 
$Q=Q_n+Q_p=\sqrt{\frac{16\pi}5} (\frac{3}{4\pi} AR^2\beta_2)$. The 
root mean square (rms) matter radius is defined as
$\langle r_m^2\rangle={1\over{A}}\int\rho(r_{\perp},z) r^2d\tau$, where
$A$ is the mass number, and $\rho(r_{\perp},z)$ is the deformed density.
The total binding energy and other observables are also obtained by using
the standard relations, given in \cite{ring90,ring90a}. We have used the 
recently proposed parameter set NL3* \cite{lal09}, which improves the 
description of the ground state properties of many nuclei over parameter 
set NL3 \cite{lala97}, and simultaneously provides an excellent description 
of excited states with collective character in spherical as well as in 
deformed nuclei. Just to compare, we have also used the parameter set
NL3 \cite{lala97}, which has been used in the last ten years with great 
success to describe many ground state properties of finite nuclei all 
over the periodic table \cite{gl96,af00,af03,af05,v95,v97,v01,v012}. 
However, in the mean time, several other relativistic mean-field 
interactions have been developed. In particular, the density dependent 
meson-exchange DD-ME1 \cite{ni02} and DD-ME2 \cite{la05} effective 
interactions. These effective interactions have been adjusted to improve 
the isovector channel, which has been the weak point of the NL3 
\cite{lala97} effective interaction, and provide a very successful 
description of different aspects of finite nuclei~\cite{v05,pa07}. 
Recently, a new DD-PC1~\cite{ni08} (density dependent point coupling) 
effective interaction has been developed, which works very well in the 
region of deformed heavy nuclei. As outputs, we obtain different 
potentials, densities, single-particle energy levels, radii, deformations 
and the binding energies. For a given nucleus, the maximum binding 
energy corresponds to the ground state and other solutions are obtained 
as various excited intrinsic states.

\subsection{Pairing Calculation}
It is well know that pairing correlations have to be included in 
any realistic calculation of medium and heavy nuclei. In principle, 
the microscopic Hartree-Fock-Bogoliubov (HFB) theory should be used, 
which have been discussed in several articles 
\cite{v05,pa07,ni08,go96,pr96,jm06,la99,la991} and in the references 
given there. However, for pairing calculations of a broad range of 
nuclei not too far from the $\beta$-stability line, a simpler approach, 
the constant gap, BCS-pairing approach is reasonably well. But, this 
simple approach breaks down for nuclei far from the valley of 
$\beta$-stability, where the coupling to the continuum is important 
\cite{do84}. In the present study, we treat the pairing correlations 
using the BCS approach. Although the BCS approach may fail for light 
neutron rich nuclei, the nuclei considered here are not light neutron 
rich nuclei and the RMF results with BCS treatment should be reliable.

The contribution of the pairing interaction to the total energy, for 
each nucleon, is
\begin{equation}
E_{pair}=-G\left[\sum_{i>0}u_{i}v_{i}\right]^2
\end{equation}
where $v_i^2$ and $u_i^2=1-v_i^2$ are the occupation probabilities, 
and $G$ is the pairing force constant~\cite{ring90,sk93,map82}. The 
variational procedure with respect to the occupation numbers $v_i^2$, 
gives the BCS equation
\begin{equation}
2\epsilon_iu_iv_i-\triangle(u_i^2-v_i^2)=0
\end{equation}
and the gap $\Delta$ is defined by
\begin{equation}\triangle=G\sum_{i>0}u_{i}v_{i}
\end{equation}
This is the famous BCS equation for pairing energy.
The densities are contained within the occupation number
\begin{equation}
n_i=v_i^2=\frac{1}{2}\left[1-\frac{\epsilon_i-\lambda}
{\sqrt{(\epsilon_i-\lambda)^2+\triangle^2}}\right]
\end{equation}
For the pairing gaps for proton and neutron, we choose the standard 
expressions, which are valid for nuclei both on or away from the 
stability line, and are given by the expressions~\cite{madland81,madland81a}:
\begin{eqnarray}
\triangle_p &=& RB_s e^{sI-tI^2}/Z^{1/3}\;\;\;\;\;\mbox{and}\\
\triangle_n &=& RB_s e^{-sI-tI^2}/A^{1/3}
\end{eqnarray}
The inputs of pairing gaps i.e., $R$ = 5.72, $s$ = 0.118, $t$ = 8.12, 
$B_s$ = 1, and $I = (N-Z)/(N+Z)$ are used in nuclear physics for many 
years. We consider that it is suitable here. The occupation probability
is calculated using Eqs. (12) and (13), and the chemical potentials
$\lambda_n$ and $\lambda_p$ are determined by the particle  numbers 
for protons and neutrons. Using Eqs. (10) and (11) the pairing energy 
for nucleon can be written as
\begin{equation}
E_{pair}= -\frac{\Delta^2}{G} =-\triangle\sum_{i>0}u_{i}v_{i}
\end{equation}
It can be seen from Eq. (14), that in the present approach to include 
the pairing effects using the constant pairing gap, the pairing energy 
$E_{pair}$ is not constant since it depends on the occupation 
probabilities $v_i^2$ and $u_i^2$, and hence on the deformation parameter 
$\beta_2$, particularly near the Fermi surface. It is known to us that 
the pairing energy $E_{pair}$ diverges if it is extended to an infinite 
configuration space for a constant pairing parameter $\triangle$ and 
force constant $G$. Also, for the states spherical or deformed, with 
large momenta near the Fermi surface, $\triangle$ decreases in all the 
realistic calculations with finite range forces. However, for the sake 
of simplicity of the calculation, we have taken constant pairing gap 
by assuming that the pairing gap for all states are equal to each other 
near the Fermi surface. In the present calculations we have used 
a pairing window, and all the equations extended up to the level 
$\epsilon_i-\lambda\leq 2(41A^{1/3})$, where a factor of 2 has been 
included in order to reproduce the pairing correlation energy for 
neutrons in $^{118}$Sn using Gogny force \cite{sk93}. This kind 
of approach to treat the pairing correlation, has already been used by 
us and many other authors in RMF model as well in non relativistic 
SHF  model \cite{bhu09,ring90,sk93,pac,apac,bpac,bh11}.
\begin{figure}
\begin{center}
\includegraphics[width=1.0\columnwidth]{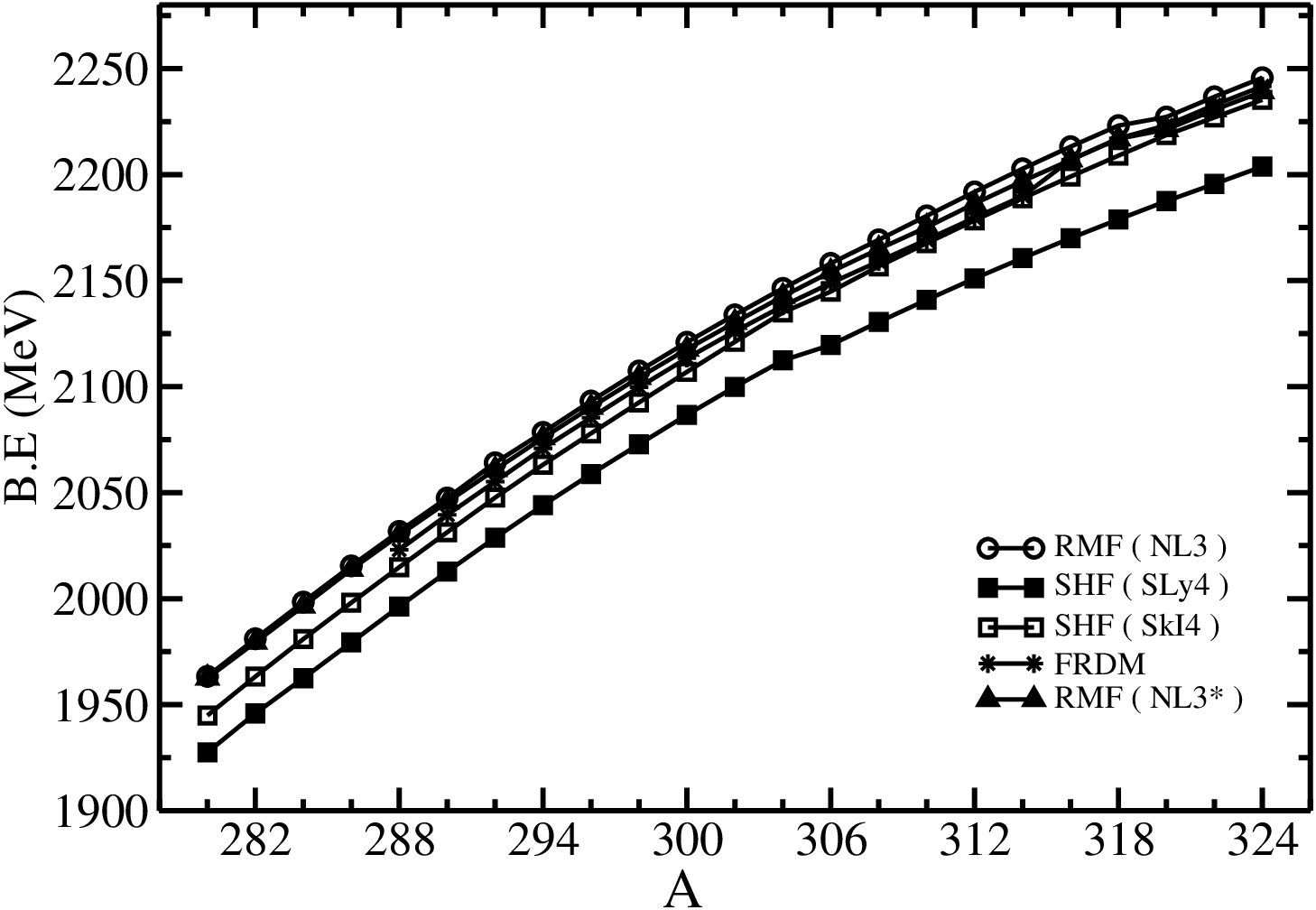}
\caption{The total binding energy BE for $^{280-324}$120 nuclei
in RMF(NL3*), RMF(NL3), SHF(SkI4) and SHF(SLy4) calculations 
compared with the FRDM results [62] wherever available.
}
\end{center}
\end{figure}

\section{Results and Discussion}
\subsection{Ground state properties using the SHF and RMF models}

There exist a number of parameter sets for solving the standard SHF
Hamiltonians and RMF Lagrangians. In many of our previous works and
those of other~\cite{patra1,ring90,lala97,patra2,patra3,patra4}
the ground state properties, like the binding energies (BE),
quadrupole deformation parameters $\beta_2$, charge radii ($r_c$),
and other bulk properties , are evaluated by using the various
nonrelativistic and relativistic parameter sets. It is found that,
more or less, most of the recent parameter sets reproduce well
the ground state properties, not only of stable normal nuclei
but also of exotic nuclei far away from the valley of 
$\beta$-stability. This means that if one uses a reasonably 
acceptable parameter set, the predictions of the model will remain 
nearly force independent. In this paper we have used the improved 
version of NL3 parameter set (NL3*), standard NL3, SkI4 and SLy4 
parameter sets for our calculations.

\begin{figure}
\begin{center}
\includegraphics[width=1.0\columnwidth]{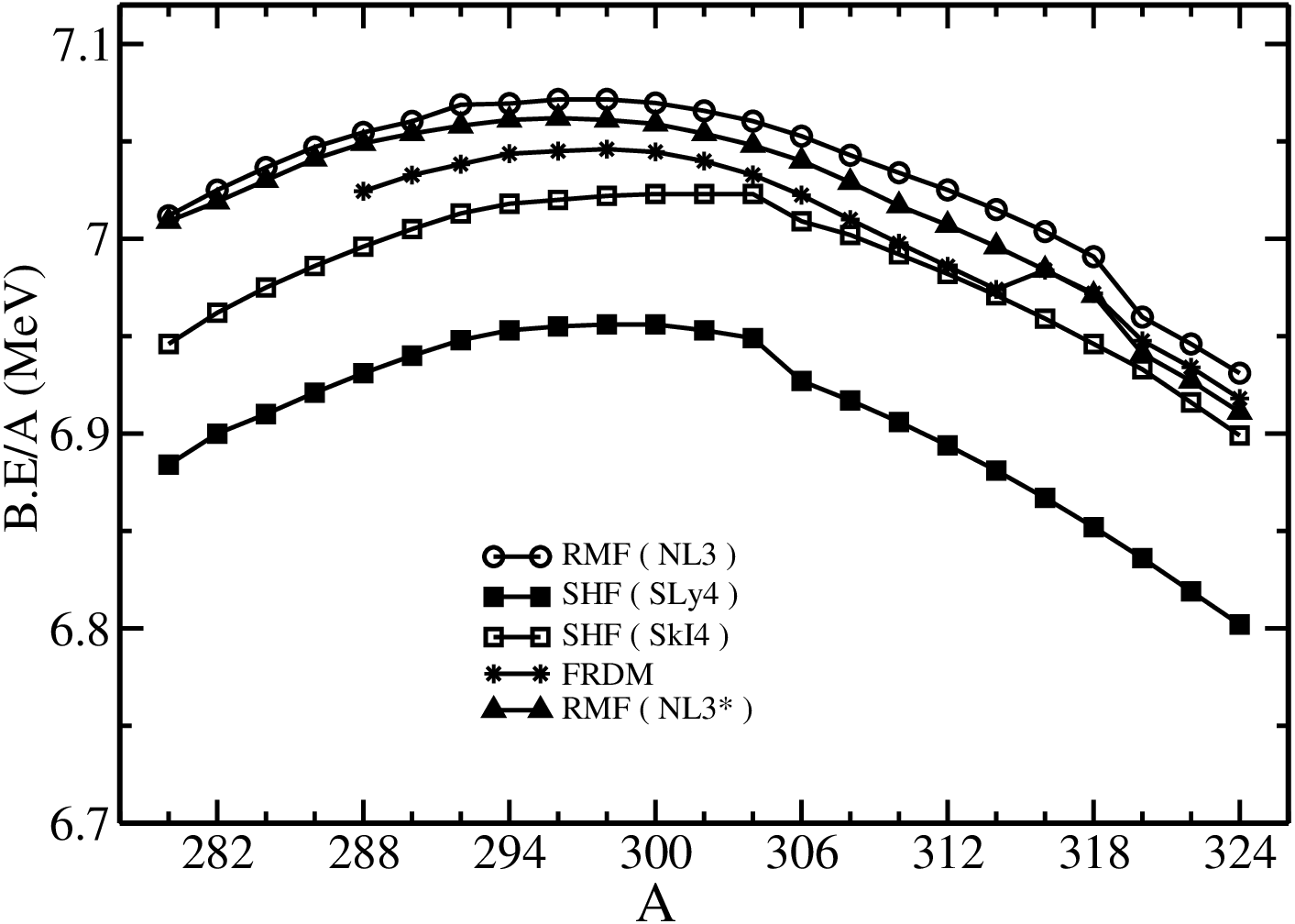}
\caption{The binding energy per particle BE/A for the superheavy
isotopes $^{280-324}$120 obtained in RMF(NL3*), RMF(NL3), SHF(SkI4) 
and SHF(SLy4) formalism compared with the FRDM results [62] wherever 
available.
}
\end{center}
\end{figure}

\subsection{Binding Energy and Two-neutron Separation Energy and 
Pairing Energy}

Binding energies are important quantities of nuclei and they are directly
related to the stability of nuclei and to $\alpha$-decay energies.
Whether a model can quantitatively reproduce the experimental binding
energy is a crucial criterion to judge the validity of the model for
superheavy nuclei. Figure 1 shows the total binding energy BE, obtained
in both nonrelativistic SHF and relativistic
RMF formalism compared with the Finite Range Droplet Model (FRDM)
results \cite{moll97,moll97a}. From the figure it is clear that, the binding
energy obtained in both the RMF and SHF models are qualitatively
similar. We notice that the binding energy, obtained using the
NL3 and NL3* parameter set are almost equal within lower mass region but,
towards higher mass region, the binding energy using NL3* parameter set,
is gradually getting lower values than NL3 parameter set, which are 
over-estimate to both the SHF (SkI4) and SHF (SLy4) results by almost 
a constant factor. For the total binding energy of the isotopic
chain in Tables 1, 2 and Figure 1, we notice that the macro-microscopic 
FRDM calculation lies in between microscopic RMF and SHF. In case of SHF 
(SkI4) the difference decreases gradually towards the higher mass region. 

\begin{figure}
\begin{center}
\includegraphics[width=1.0\columnwidth]{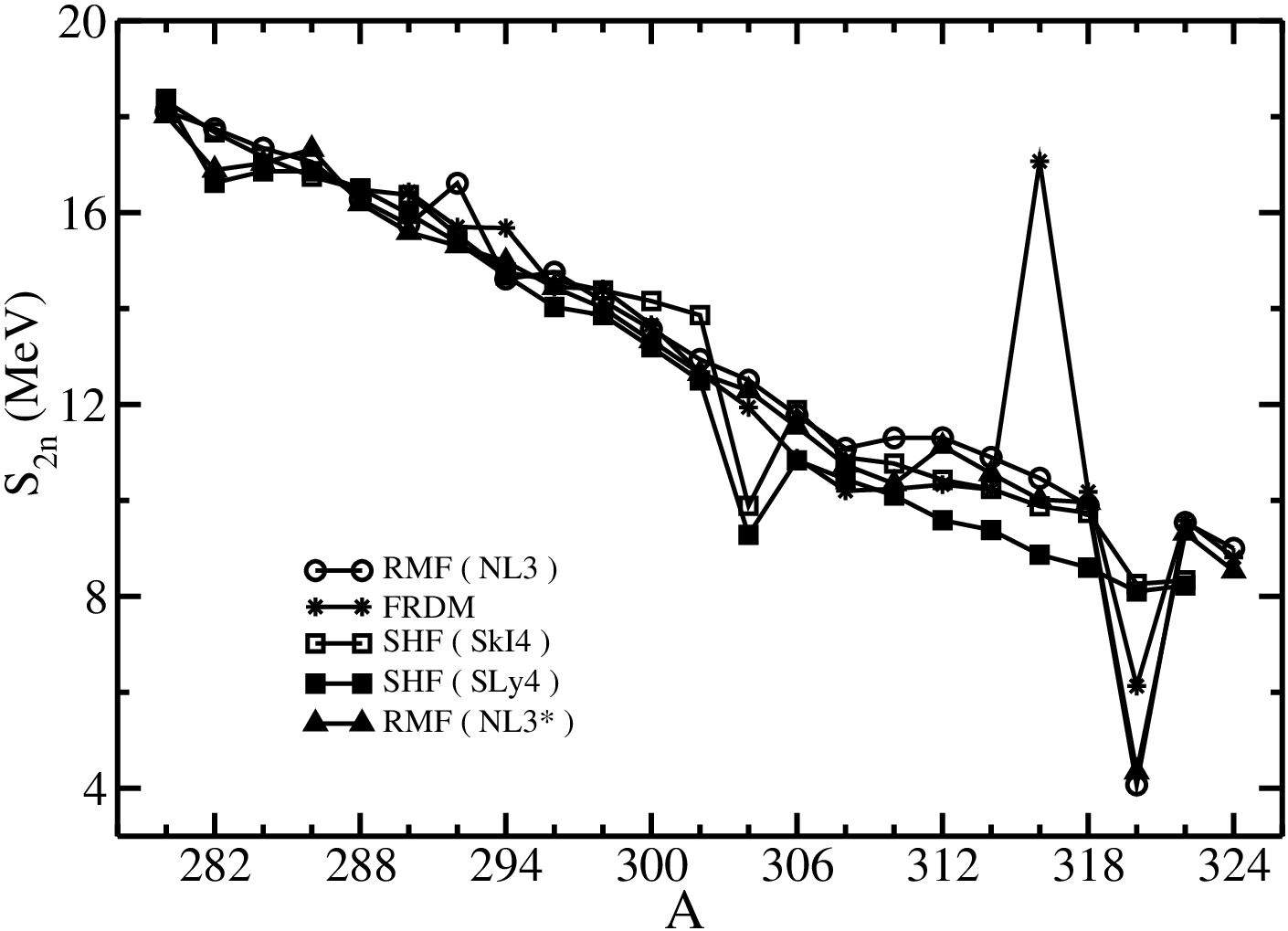}
\caption{The two-neutron separation energy $S_{2n}$ for $^{280-324}$120 
nuclei, obtained from RMF(NL3*), RMF(NL3), SHF(SkI4) and SHF(SLy4) 
formalisms and compared with the FRDM results [62] wherever available.
}
\end{center}
\end{figure}

The binding energy per particle (BE/A) for the isotopic chain 
is also plotted in Figure 2. We notice that the SHF and RMF curves 
could almost be overlapped with one another through a constant 
scaling factor. The FRDM calculation lies in between RMF and SHF. 
This means, qualitatively, all the curves show a similar behavior. 
In general, the BE/A value starts increasing with the increase of 
mass number A, reaching a peak value at A $\sim$ 304 for RMF, 
SHF and FRDM models. This means that $^{304}120$ is the most stable
element from the binding energy point of view, which is situated 
at A$\sim$ 304 (N=184, Z=120). Interestingly, this neutron number 
are close to N = 184, which is the next predicted magic number 
\cite{bunu12}. It is worthy to mention that, the results obtained 
in the present calculation are almost consistents to the prediction 
by earlier calculations \cite{rutz97,rutz99,ren001,ren002,ren03,ren04,zha05} 
using some different force parameters. Hence, we may note that 
the results for binding energy and related observables like magic 
numbers are independent of force parameters. 

In Tables 1 and 2, we also show a comparison of the calculated 
two-neutron separation energy $S_{2n}$(N,Z) = BE(N,Z)-BE(N-2,Z) 
with the Finite Range Droplet Model (FRDM) predictions of Ref. 
\cite{moll97,moll97a}, wherever possible. From the tables, we find that
the microscopic $S_{2n}$ values agree well with the macro-microscopic
FRDM calculations. The comparisons of $S_{2n}$ for RMF and SHF models 
with the FRDM result are further shown in Figure 3, which clearly shows 
that the RMF and the FRDM $S_{2n}$ values coincide remarkably well, 
except at mass A = 316 which seems spurious due to some error somewhere 
in the case of FRDM. Apparently, the $S_{2n}$ decrease gradually with 
increase of neutron number, except for the noticeable kinks at
A = 282 (N = 172) and 318 (N = 198) in RMF(NL3*), at A = 292 (N = 172) 
and 318 (N = 198) in RMF(NL3), at A = 294 (N = 174) and 318 (N = 198) 
in FRDM, at A = 282 (N = 162) and 304 (N = 184) in SHF (SLy4) and
at A = 304 (N = 184) and 318 (N = 198) in SHF (SkI4).
Interestingly, these neutron numbers are close to either 
N = 172 or 184 magic numbers.

\begin{figure}
\begin{center}
\includegraphics[width=1.0\columnwidth]{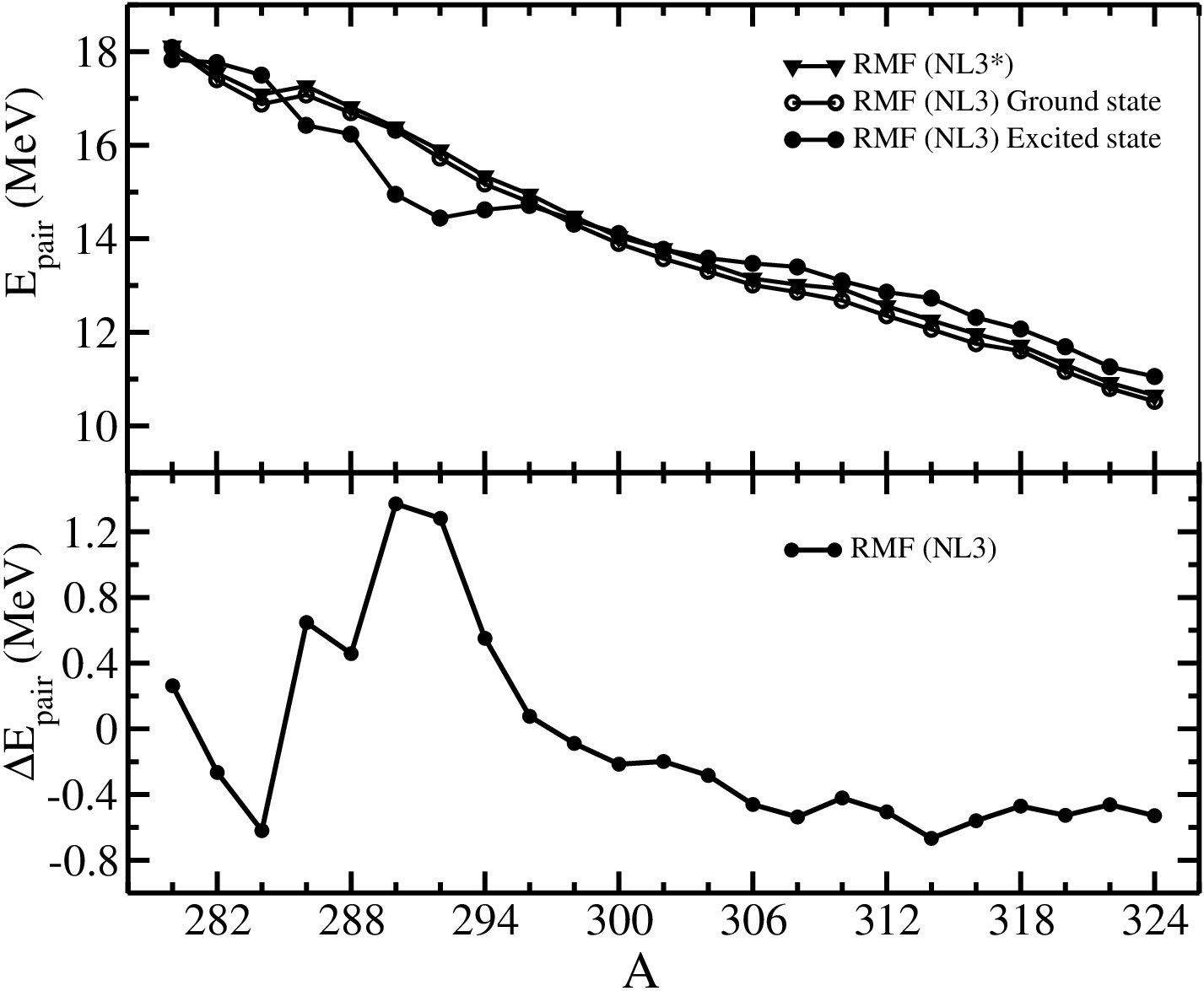}
\caption{(a) The Pairing energy $E_{pair}$ for the ground state 
and first excited states and (b) the pairing energy difference 
$\Delta E_{pair}$ between the $E_{pair}$ values of ground and 
first excited states [$\Delta E_{pair}$ = $E_{pair}$(g.s.) 
- $E_{pair}$(e.s.)] for $^{280-324}$120 nuclei as a function of 
mass number, using the RMF formalism with NL3* and  NL3 parameter 
set.}
\end{center}
\end{figure}

Figure 4, show the pairing energy $E_{pair}$ as well as the 
pairing energy difference $\Delta E_{pair}$ as a function of 
mass number A. In Figure 4(a), we show the pairing energy 
$E_{pair}$ for both the ground state (g.s.) using RMF(NL3*) 
and RMF(NL3), and the first excited state (e.s.) using RMF(NL3), 
referring to different $\beta_2$ values for the full isotopic 
chain. The difference in the two $E_{pair}$ values, i.e.
$\Delta E_{pair}$ = $E_{pair}$(g.s.) - $E_{pair}$(e.s.),
is shown in Figure 4(b). It is clear from Figure 4(a) that
$E_{pair}$ decreases with an increase in mass number A; i.e., 
even if the $\beta_2$ values for two nuclei are the same, the 
pairing energies are different from one another. From the above 
results, it can be seen that for a given nucleus, pairing energy  
$E_{pair}$ depends only marginally on the quadrupole deformation 
$\beta_2$. On the other hand, even if the $\beta_2$ values for 
two nuclei are same, the $E_{pair}$ values are different from one
another, depending on the filling of the nucleons.

\subsection{Shape coexistence}

\begin{figure}[ht]
\begin{center}
\includegraphics[width=1.0\columnwidth]{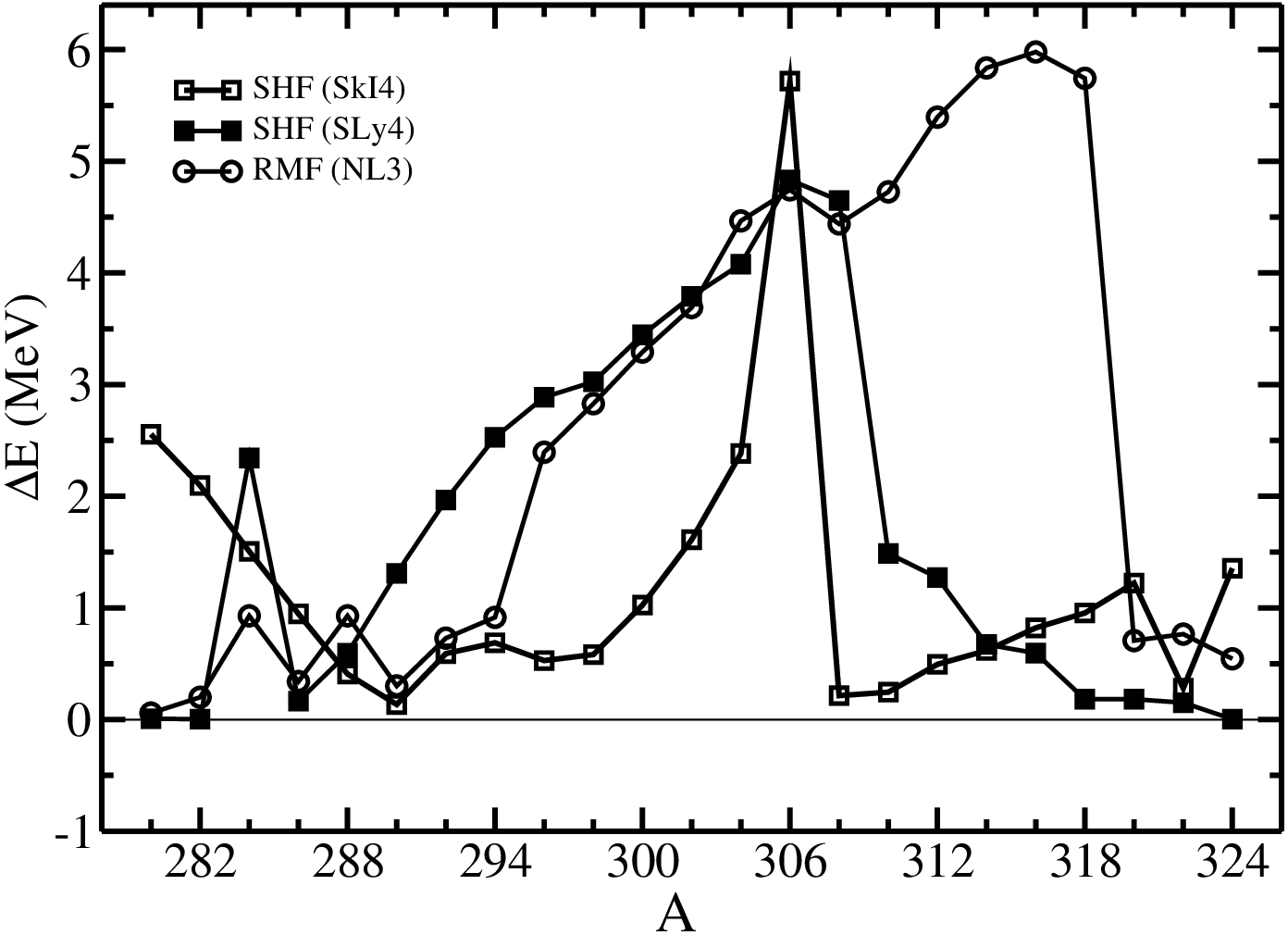}
\caption{The energy difference between the ground state and the first
excited state in both nonrelativistic SHF(SkI4), SHF(SLy4) and
RMF formalisms with NL3* and NL3 parameter set.
}
\end{center}
\end{figure}

We have also calculated the existing other solutions for the whole
Z = 120 isotopic chain, both in prolate and oblate deformed
configurations. In many cases, we find low-lying excited states.
As a measure of the energy difference between the ground
band and the first excited state, we have plotted in Figure 5
the binding energy difference $\triangle E$ between
the two solutions, noting that the maximum binding energy solution
refers to the ground state (g.s.) and all other
solutions to the intrinsic excited states (e.s.). From Figure 5,
we notice that, in RMF calculations, the energy difference
$\triangle E$ is small for neutron-deficient isotopes,
but it increases with increase of mass number A in the isotopic
series. This small difference in the binding energy
for neutron-deficient isotopes is an indication
of shape coexistence. In other words, the two solutions
in these nuclei are almost degenerate for a small difference of
output in energy. For example, in $^{290}$120, the two
solutions for $\beta_2$ = 0.555 and $\beta_2$ = 0.002 are completely
degenerate with binding energies of 2047.503 and 2047.202 MeV.
This later result suggests that the ground state can be changed
to the excited state and vice-versa by a small change
in the input, like the pairing strength, etc., in the calculations.
Similarly, in case of SHF (SkI4) calculation,
the energy difference  $\triangle E$
remains small between mass number A = 286 to 300 and between mass
number A = 308 to 324, indicating of the presence of shape
coexistence. We also have the indication of the presence of
shape coexistence through SHF (SLy4) calculation.
In any case, such a phenomenon is known to exist
in many other regions of the periodic table \cite{patra94,patra94a,patra94b,patra94c}.

\begin{table}
\caption{The RMF(NL3*), RMF(NL3), SHF (SkI4) and SHF (SLy4) results
for binding energy BE, the quadrupole deformation parameter $\beta_{2}$,
two-neutron separation energy $S_{2n}$ and the binding energy difference
$\triangle E$ between the ground- and first-exited state solutions, compared
with the corresponding Finite Range Droplet Model (FRDM) results [62] for 
the isotopic chain of $^{280-302}$120. The energy is in MeV.
}
\renewcommand{\tabcolsep}{0.4cm}
\renewcommand{\arraystretch}{0.6}
\begin{tabular}{cccccc}
\hline
Nucleus & Formalism & BE & $\beta_{2}$ & $S_{2n}$ & $\Delta E$ \\
\hline
280 & RMF (NL3) & 1963.33  & 0.258 & 18.11 & 0.058 \\
& RMF (NL3*)& 1962.54  & 0.258 & 18.02 &      \\
& SHF (SkI4)& 1944.92  & 0.529 & 18.32 & 2.555 \\
& SHF (SLy4)& 1927.51  & 0.227 & 18.37 & 0.008 \\
282 & RMF (NL3) & 1981.08  & -0.422& 17.75 & 0.200 \\
& RMF (NL3*)& 1979.43  & -0.426& 16.89 &      \\
& SHF (SkI4)& 1963.24  & 0.520 & 17.68 & 2.096  \\
& SHF (SLy4)& 1945.88  & 0.227 & 16.62 & 0.003  \\
284 & RMF (NL3) & 1998.42  & -0.428& 17.34 & 0.929 \\
& RMF (NL3*)& 1996.46  & -0.432& 17.03 &      \\
& SHF (SkI4)& 1980.91  & 0.519 & 17.16 & 1.507 \\
& SHF (SLy4)& 1962.51  & 0.202 & 16.86 & 2.343 \\
286 & RMF (NL3) & 2015.48  & 0.567 & 17.16 & 0.340 \\
& RMF (NL3*)& 2013.78  & 0.567 & 17.32 &      \\
& SHF (SkI4)& 1998.08  & 0.521 & 16.76 & 0.947 \\
& SHF (SLy4)& 1979.37  & 0.537 & 16.86 & 0.165 \\
288 & RMF (NL3) & 2031.75  & 0.560 & 16.28 & 0.929 \\
& RMF (NL3*)& 2029.97  & 0.562 & 16.19 &      \\
& SHF (SkI4)& 2014.83  & 0.523 & 16.48 & 0.408 \\
& SHF (SLy4)& 1996.23  & 0.122 & 16.51 & 0.593 \\
& FRDM      & 2023.03  & -0.113&       &       \\
290 & RMF (NL3) & 2047.50  & 0.551 & 15.75 & 0.301 \\
& RMF (NL3*)& 2045.56  & 0.556 & 15.59 &      \\
& SHF (SkI4)& 2031.31  & 0.119 & 16.37 & 0.135 \\
& SHF (SLy4)& 2012.74  & 0.115 & 15.98 & 1.310 \\
& FRDM      & 2039.49  &       &       &       \\
292 & RMF (NL3) & 2064.11  & 0.540 & 16.61 & 0.730 \\
& RMF (NL3*)& 2060.87  & 0.547 & 15.31 &      \\
& SHF (SkI4)& 2047.68  & 0.113 & 15.53 & 0.591 \\
& SHF (SLy4)& 2028.71  & 0.107 & 15.38 & 1.966 \\
& FRDM      & 2055.19  & -0.130& 15.70 &       \\
294 & RMF (NL3) & 2078.43  & 0.536 & 14.61 & 0.916 \\
& RMF (NL3*)& 2075.85  & 0.541 & 14.98 &      \\
& SHF (SkI4)& 2063.21  & 0.110 & 14.75 & 0.688 \\
& SHF (SLy4)& 2044.09  & 0.097 & 14.69 & 2.528 \\
& FRDM      & 2070.87  & 0.081 & 15.68 &       \\
296 & RMF (NL3) & 2093.19  & 0.542 & 14.76 & 2.394 \\
& RMF (NL3*)& 2090.29  & 0.545 & 14.44 &      \\
& SHF (SkI4)& 2077.96  & 0.087 & 14.59 & 0.529 \\
& SHF (SLy4)& 2058.78  & 0.088 & 14.03 & 2.887 \\
& FRDM      & 2085.32  & -0.096& 14.45 &       \\
298 & RMF (NL3) & 2107.35  & 0.551 & 14.16 & 0.058 \\
& RMF (NL3*)& 2104.30  & 0.554 & 14.01 &      \\
& SHF (SkI4)& 2092.55  & 0.066 & 14.38 & 0.583 \\
& SHF (SLy4)& 2072.81  & 0.060 & 13.86 & 3.026 \\
& FRDM      & 2099.73  & -0.079& 14.41 &       \\
300 & RMF (NL3) & 2120.92  & 0.561 & 13.57 & 3.292 \\
& RMF (NL3*)& 2117.63  & 0.564 & 13.33 &      \\
& SHF (SkI4)& 2106.94  & 0.045 & 14.16 & 1.026 \\
& SHF (SLy4)& 2086.68  & 0.040 & 13.19 & 3.446 \\
& FRDM      & 2113.39  & -0.008& 13.66 &       \\
302 & RMF (NL3) & 2133.86  & 0.579 & 12.94 & 3.691 \\
& RMF (NL3*)& 2130.28  & 0.586 & 12.65 &      \\
& SHF (SkI4)& 2121.09  & 0.024 & 13.86 & 1.611 \\
& SHF (SLy4)& 2099.87  & 0.019 & 12.5  & 3.791 \\
& FRDM      & 2126.05  & 0.000 & 12.66 &       \\
\hline
\end{tabular}
\label{Table 1}
\end{table}

\subsection{Quadrupole deformation parameter}

\begin{figure}
\begin{center}
\includegraphics[width=1.0\columnwidth]{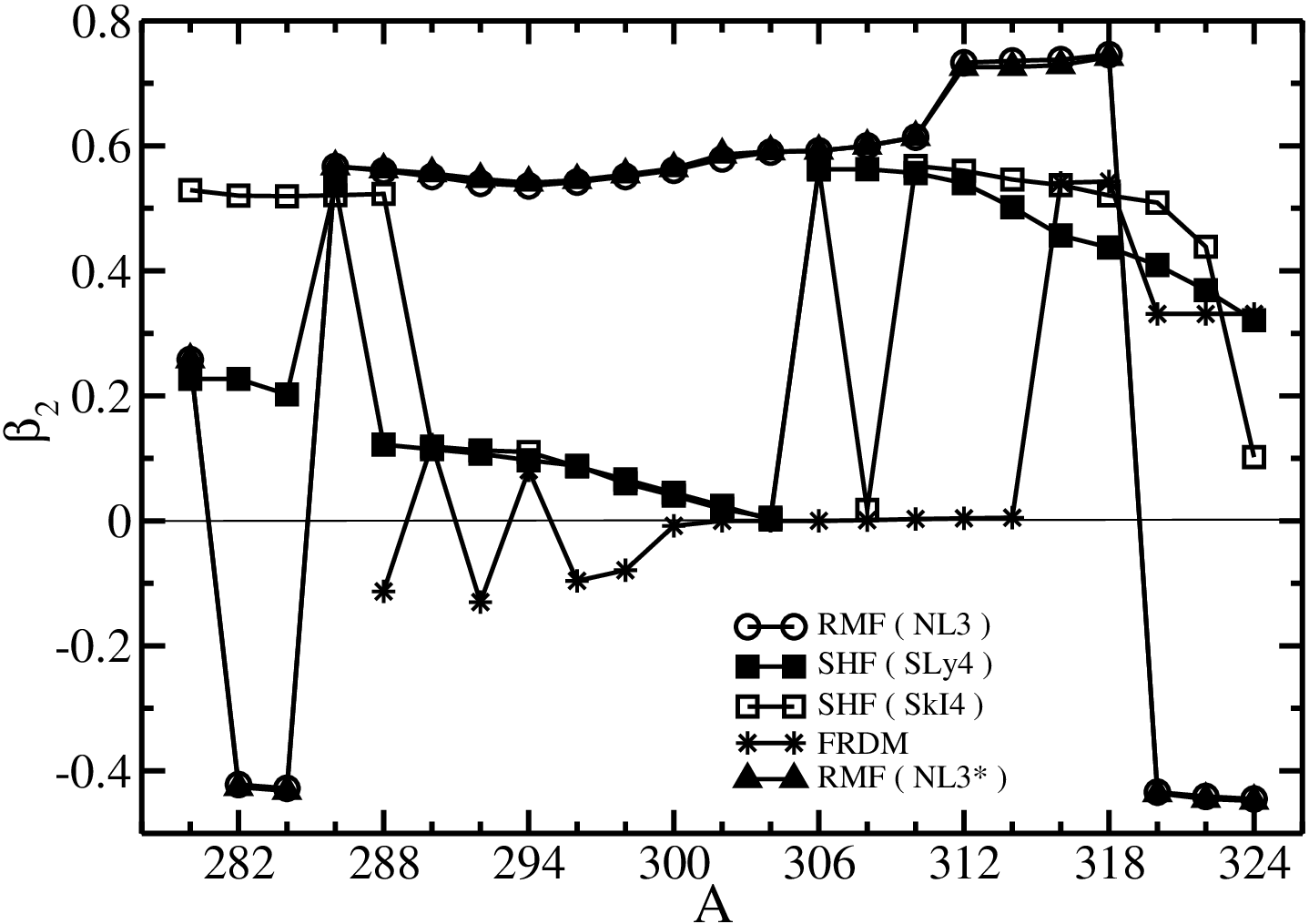}
\caption{Comparison of quadrupole deformation parameter obtained
from nonrelativistic SHF(SkI4), SHF(SLy4) and relativistic mean-field 
formalism (RMF) with NL3* and NL3 parameter set, compared with the FRDM 
results [62] wherever available.
}
\end{center}
\end{figure}

The quadrupole deformation parameter $\beta_2$, for both the
ground and first excited states, is also determined within
the two formalisms. In some of the earlier RMF and SHF calculations,
it was shown that the quadrupole
moment obtained from these theories reproduce the experimental data
pretty well \cite{patra1,cha97,cha98,sero86,ring90,lala97,patra2,rei95,brown98}.
The ground state (g.s.) quadrupole deformation parameter
$\beta_2$ values plotted in Figure 6 for SHF and RMF formalisms,
and compared with the FRDM results \cite{moll97,moll97a} show that
the FRDM results differ strongly along the whole mass regions.
In both the SHF (SkI4) and SHF (SLy4) results, we find that
the solutions for the whole isotopic chain are prolate. However,
in some mass region, we find highly deformed prolate solutions
in the ground state configuration.
In RMF formalism using both NL3* and NL3 parameter set,
 we find shape change from prolate to highly deformed
oblate at A = 282. Then, with increase in mass number there is
a shape change from highly oblate to highly prolate, and
again we find a shape change from highly prolate to highly
oblate at A = 320. A more careful inspection show that the
solutions for the whole isotopic chain are prolate, except
at A = 282, 284 and at A = 320-324 for both the RMF(NL3*)
and RMF(NL3) model.
Interestingly, most of the isotopes are superdeformed in their
ground state configurations, and because of the shape
coexistence properties of these isotopes, sometimes it is
possible that the ground state could be the near spherical
solution.

\begin{table}
\caption{Same as TABLE I, for the isotopic chain
of $^{304-324}$120.
}
\renewcommand{\tabcolsep}{0.4cm}
\renewcommand{\arraystretch}{0.6}
\begin{tabular}{ccccccccc}
\hline
\hline
Nucleus & Formalism & BE & $\beta_{2}$ & $S_{2n}$ & $\Delta E$ \\
\hline
304 & RMF (NL3) & 2146.37  & 0.590 & 12.51 & 4.466 \\
    & RMF (NL3*)& 2142.57  & 0.591 & 12.29 &      \\
    & SHF (SkI4)& 2134.96  & 0.002 & 9.89  & 2.383 \\
    & SHF (SLy4)& 2112.37  & 0.005 & 9.38  & 4.078 \\
    & FRDM      & 2137.99  & 0.000 & 11.94 &       \\
306 & RMF (NL3) & 2158.14  & 0.592 & 11.78 & 4.744 \\
    & RMF (NL3*)& 2154.10  & 0.596 & 11.53 &      \\
    & SHF (SkI4)& 2144.85  & 0.564 & 11.88 & 5.718 \\
    & SHF (SLy4)& 2119.65  & 0.562 & 10.84 & 4.833 \\
    & FRDM      & 2148.87  & 0.000 & 10.88 &       \\
308 & RMF (NL3) & 2169.23  & 0.600 & 11.09 & 4.439 \\
    & RMF (NL3*)& 2164.84  & 0.600 & 10.74 &      \\
    & SHF (SkI4)& 2156.73  & 0.018 & 10.90 & 0.215 \\
    & SHF (SLy4)& 2130.49  & 0.563 & 10.44 & 4.649 \\
    & FRDM      & 2159.06  & 0.001 & 10.20 &       \\
310 & RMF (NL3) & 2180.54  & 0.614 & 11.30 & 4.727 \\
    & RMF (NL3*)& 2175.19  & 0.614 & 10.35 &      \\
    & SHF (SkI4)& 2167.62  & 0.568 & 10.77 & 0.245 \\
    & SHF (SLy4)& 2140.93  & 0.557 & 10.10 & 1.488 \\
    & FRDM      & 2169.30  & 0.003 & 10.24 &       \\
312 & RMF (NL3) & 2191.84  & 0.733 & 11.30 & 5.396 \\
    & RMF (NL3*)& 2186.32  & 0.726 & 11.13 &      \\
    & SHF (SkI4)& 2178.39  & 0.560 & 10.43 & 0.495 \\
    & SHF (SLy4)& 2151.03  & 0.540 & 9.59  & 1.272 \\
    & FRDM      & 2179.63  & 0.004 & 10.33 &       \\
314 & RMF (NL3) & 2202.74  & 0.736 & 10.89 & 5.837 \\
    & RMF (NL3*)& 2196.88  & 0.726 & 10.56 &      \\
    & SHF (SkI4)& 2188.82  & 0.546 & 10.24 & 0.618 \\
    & SHF (SLy4)& 2160.62  & 0.502 & 9.38  & 0.672 \\
    & FRDM      & 2189.85  & 0.005 & 10.23 &       \\
316 & RMF (NL3) & 2213.20  & 0.733 & 10.46 & 5.979 \\
    & RMF (NL3*)& 2206.90  & 0.729 & 10.02 &      \\
    & SHF (SkI4)& 2199.06  & 0.537 & 9.88  & 0.821 \\
    & SHF (SLy4)& 2170.01  & 0.457 & 8.87  & 0.595 \\
    & FRDM      & 2206.93  & 0.541 & 17.07 &       \\
318 & RMF (NL3) & 2223.10  & 0.746 & 9.89  & 5.743 \\
    & RMF (NL3*)& 2216.86  & 0.742 & 9.96  &      \\
    & SHF (SkI4)& 2208.94  & 0.521 & 9.74  & 0.955 \\
    & SHF (SLy4)& 2178.88  & 0.437 & 8.60  & 0.183 \\
    & FRDM      & 2217.10  & 0.543 & 10.18 &       \\
320 & RMF (NL3) & 2227.16  & -0.434& 4.07  & 0.707 \\
    & RMF (NL3*)& 2221.24  & -0.436& 4.34  &      \\
    & SHF (SkI4)& 2218.68  & 0.509 & 8.256 & 1.223 \\
    & SHF (SLy4)& 2187.48  & 0.409 & 8.103 & 0.183 \\
    & FRDM      & 2223.24  & 0.331 & 6.13  &       \\
322 & RMF (NL3) & 2236.71  & -0.441& 9.54  & 0.765 \\
    & RMF (NL3*)& 2230.56  & -0.445& 9.32  &      \\
    & SHF (SkI4)& 2226.94  & 0.439 & 8.33  & 0.280 \\
    & SHF (SLy4)& 2195.58  & 0.369 & 8.23  & 0.151 \\
    & FRDM      & 2232.82  & 0.331 & 9.58  &       \\
324 & RMF (NL3) & 2245.71  & -0.445& 9.00  & 0.544 \\
    & RMF (NL3*)& 2239.09  & -0.448& 8.53  &      \\
    & SHF (SkI4)& 2235.27  & 0.102 &       & 1.355 \\
    & SHF (SLy4)& 2203.81  & 0.321 &       & 0.004 \\
    & FRDM      & 2241.63  & 0.331 & 8.81  &       \\
\hline
\end{tabular}
\end{table}

\section{The $Q_{\alpha}$ energy and the decay half-life $T_{1/2}^{\alpha}$}
The $Q_{\alpha}$ energy is obtained from the relation \cite{patra23}:
[$Q_{\alpha}$ (N, Z) = BE (N, Z)-BE (N-2, Z-2)-BE (2, 2).]
Here, $BE$ (N, Z) is the binding energy of the parent nucleus
with neutron number N and proton number Z, $BE$ (2, 2) is the
binding energy of the $\alpha$-particle ($^4$He), i.e., 28.296 MeV,
and $BE$ (N-2, Z-2) is the binding energy of the
daughter nucleus after the emission of an $\alpha$-particle.

The half-life time $log_{10}T_{1/2}^{\alpha} (s)$ values are
estimated by using the phenomenological
formula of Viola and Seaborg \cite{viol01}:
\begin{equation}
log_{10}T^{\alpha}_{1/2}(s)=\frac {aZ-b}{\sqrt{Q_{\alpha}}}-(cZ+d) + h_{log},
\end{equation}
where Z is the proton number of the parent nucleus.
For the $a$, $b$, $c$ and $d$ parameters we consider the Sobiczewski {\it et al.}
modified values obtained using more recent and expanded data base of
even-even nuclides~\cite{sobi89}, which are: $a$ = 1.66175;
$b$ = 8.5166; $c$ = 0.20228; $d$ = 33.9069.
The quantity $h_{log}$ accounts for the hindrances associated
with the odd proton and neutron numbers as given by
Viola and Seaborg \cite{viol01}, namely
$h_{log}=\begin{array}{ll}
           0, &  \mbox{ Z and N even}\\
        0.772,&  \mbox{ Z odd and N even}\\ 
        1.066,&  \mbox{ Z even and N odd}\\ 
        1.114,&  \mbox{ Z and N odd}. 
        \end{array}$
\begin{table}
\caption{The $Q_{\alpha}$ and $log_{10}T_{1/2}^{\alpha}$ (in sec) for
$\alpha$-decay series of $^{292}$120 nucleus, calculated in the SHF
(SkI4), SHF (SLy4), RMF(NL3*) and RMF(NL3) models, and compared with the Finite
Range Droplet Model (FRDM) results [62] and other theoretical
results [29,67,68,69], wherever available.
In order to compare, we have calculated the $T_{1/2}^{\alpha}$
values in case of Sobiczewski {\it et al.}
using Eq.(15).
}
\renewcommand{\tabcolsep}{0.40cm}
\renewcommand{\arraystretch}{0.6}
\begin{tabular}{ccccccccc}
\hline
A & Z & Formalism & BE & $\beta_2$ & $Q_{\alpha}$ & $log_{10}T_{1/2}^{\alpha}$ \\
\hline
292&120 & RMF(NL3*)  & 2060.87 & 0.55 & 11.67&-2.36\\
   &    & RMF(NL3)   & 2064.11 & 0.54 & 10.62&0.4 \\
   &    &            & 2063.38 & 0.00 & 10.85&-0.23 \\
   &    & SHF(SKI4)  & 2047.68 & 0.11 & 12.54&-4.27 \\
   &    &            & 2047.27 & 0.53 & 13.42&-6.07 \\
   &    & SHF(SLy4)  & 2028.71 & 0.11 & 13.01&-5.65 \\
   &    &            & 2026.75 & 0.55 & 13.03&-5.26 \\
   &    & FRDM       & 2055.19 & -0.13& 13.89&-6.96 \\
288&118 & RMF(NL3*)  & 2044.27 & 0.55 & 12.41&-4.52\\
   &    & RMF(NL3)   & 2046.43 & 0.54 & 12.40&-4.51 \\
   &    &            & 2045.93 & -0.01& 11.63&-2.77     \\
   &    & SHF(SKI4)  & 2032.39 & 0.14 & 12.76&-5.27     \\
   &    &            & 2031.92 & 0.55 & 11.61&-2.74     \\
   &    & SHF(SLy4)  & 2013.43 & 0.15 & 12.82&-5.39     \\
   &    &            & 2011.47 & 0.54 & 12.47&-4.66     \\
   &    & FRDM       & 2040.79 & 0.08 & 12.87&-5.5      \\
284&116 & RMF(NL3*)  & 2028.38 & 0.18 & 11.99&-4.17\\
   &    & RMF(NL3)   & 2030.53 & 0.18 & 12.36&-4.97\\
   &    &            & 2029.26 & 0.54 & 8.54&5.68       \\
   &    & SHF(SKI4)  & 2016.85 & 0.18 & 12.5&-5.26      \\
   &    &            & 2015.71 & 0.54 & 8.99&4.07       \\
   &    & SHF(SLy4)  & 1997.94 & 0.19 & 12.19&-4.59     \\
   &    &            & 1995.64 & 0.53 & 8.75&4.92       \\
   &    & FRDM       & 2025.37 & 0.08 & 11.6&-3.28      \\
280&114 & RMF(NL3*)  & 2012.08 & 0.19 & 11.09&-2.63\\
   &    & RMF(NL3)   & 2014.59 & 0.19 & 11.14&-2.76\\
   &    &            & 2009.50 & -0.13& 9.29&2.39       \\
   &    & SHF(SKI4)  & 2001.05 & 0.93 & 13.03 &-6.85 \\
   &    &            & 1996.39 & 0.31 & 17.69&-13.95 \\
   &    & SHF(SLy4)  & 1981.83 & 0.2  & 12.32&-5.42     \\
   &    &            & 1976.09 & 0.34 & 18.06&-14.02 \\
   &    & FRDM       & 2008.67 & 0.05 & 11.61&-3.88     \\
   &    & Sobiczewski&         & 0.19 & 12.33&-5.44     \\
276&112 & RMF(NL3*)  & 1994.87 & 0.21 & 11.29&-3.72\\
   &    & RMF(NL3)   & 1997.43 & 0.21 & 11.13&-3.33 \\
   &    &            & 1990.49 & -0.15& 8.88&3.04 \\
   &    & SHF(SKI4)  & 1985.78 & 0.23 & 13.27&-7.81     \\
   &    & SHF(SLy4)  & 1965.85 & 0.23 & 12.74&-6.81     \\
   &    & FRDM       & 1991.99 & 0.21 & 11.84&-4.95     \\
   &    & Sobiczewski&         & 0.21 & 12.12&-5.55     \\
272&110 & RMF(NL3*)  & 1977.87 & 0.24 & 10.22&-1.63\\
   &    & RMF(NL3)   & 1980.26 & 0.24 & 10.61&-2.65 \\
   &    &            & 1971.07 & -0.3 & 9.84&-0.6 \\
   &    & SHF(SKI4)  & 1970.75 & 0.25 & 10.91&-3.4 \\
   &    & SHF(SLy4)  & 1950.30 & 0.25 & 10.95&-3.49     \\
   &    & FRDM       & 1975.53 & 0.22 & 10.04&-1.15     \\
   &    & Sobiczewski&         & 0.23 & 10.74&-2.98     \\
268&108 & RMF(NL3*)  & 1959.79 & 0.26 & 9.67&-0.77\\
   &    & RMF(NL3)   & 1962.57 & 0.26 & 9.66&-0.75 \\
   &    &            & 1952.61 & -0.32& 9.24&0.49       \\
   &    & SHF(SKI4)  & 1953.36 & 0.26 & 9.15&0.76 \\
   &    & SHF(SLy4)  & 1932.95 & 0.27 & 9.09&0.94 \\
   &    & FRDM       & 1957.28 & 0.23 & 9       &1.24 \\
   &    & Sobiczewski&         & 0.24 & 9.49&-0.26 \\
\hline
\end{tabular}
\end{table}

\begin{table}
\caption{Same as Table 3}
\renewcommand{\tabcolsep}{0.40cm}
\renewcommand{\arraystretch}{0.6}
\begin{tabular}{ccccccccc}
\hline
A & Z & Formalism & BE & $\beta_2$ & $Q_{\alpha}$ & $log_{10}T_{1/2}^{\alpha}$ \\
\hline
264&106 & RMF(NL3*)  & 1941.16 & 0.28 & 8.33&2.74\\
   &    & RMF(NL3)   & 1943.93 & 0.27 & 8.34&2.69 \\
   &    &            & 1933.55 & -0.31& 8.90&0.84 \\
   &    & SHF(SKI4)  & 1934.21 & 0.27 & 8.26&2.96 \\
   &    & SHF(SLy4)  & 1913.74 & 0.28 & 9.25&-0.24 \\
   &    & FRDM       & 1937.98 & 0.23 & 8.70&1.47 \\
   &    & Sobiczewski&         & 0.25 & 8.94&0.71 \\
260&104 & RMF(NL3*)  & 1921.19 & 0.28 & 7.56&4.83\\
   &    & RMF(NL3)   & 1923.97 & 0.28 & 7.75&4.08 \\
   &    &            & 1914.15 & -0.3 & 8.8 &0.44 \\
   &    & SHF(SKI4)  & 1914.18 & 0.28 & 8.46&1.55 \\
   &    & SHF(SLy4)  & 1894.69 & 0.29 & 8.46&1.54 \\
   &    & FRDM       & 1918.39 & 0.23 & 8.96&-0.05 \\
   &    & Sobiczewski&         & 0.25 & 8.84&0.32 \\
256&102 & RMF(NL3*)  & 1900.45 & 0.29 & 6.99&6.37\\
   &    & RMF(NL3)   & 1903.42 & 0.28 & 7.11&5.87 \\
   &    &            & 1894.65 & -0.2 & 7.89&2.77 \\
   &    & SHF(SKI4)  & 1894.34 & 0.3  &     &     \\
   &    & SHF(SLy4)  & 1874.85 & 0.3  &     &     \\
   &    & FRDM       & 1899.05 & 0.24 & 8.57&0.44 \\
   &    & Sobiczewski&         & 0.25 & 8.36&1.14 \\
\hline
\end{tabular}
\end{table}

\subsection{The $\alpha$-decay series of $^{292}$120 nucleus}
\begin{figure}[t]
\begin{center}
\includegraphics[width=1.0\columnwidth]{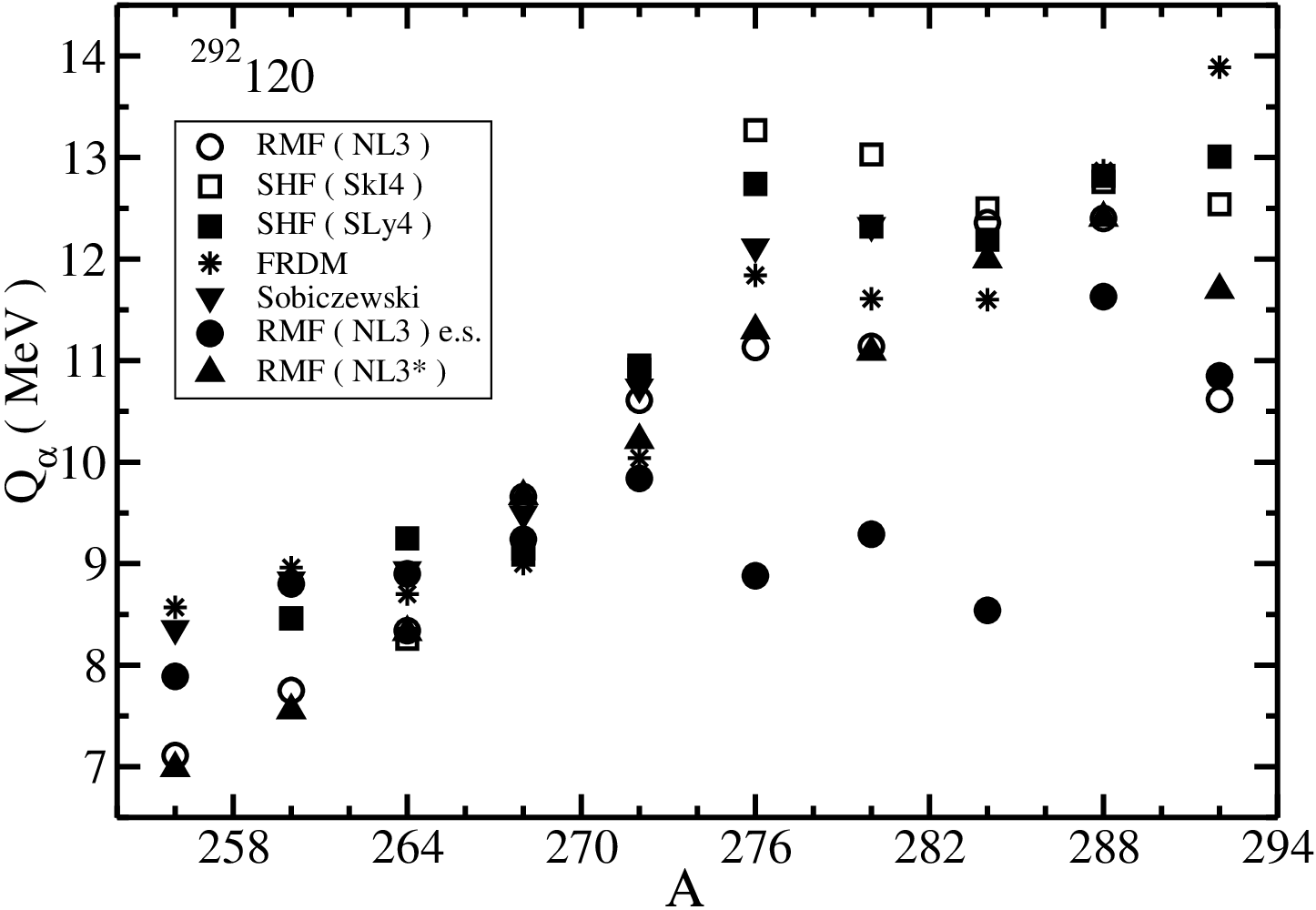}
\caption{The $Q_\alpha$ energy for the $\alpha$-decay
chain of the $^{292}$120 nucleus using nonrelativistic
SHF (SkI4), SHF (SLy4) and relativistic mean-field formalism
(RMF) with NL3* and NL3 parameter set, compared with the 
FRDM [62] and other theoretical results [29,67,68,69] wherever 
available.}
\end{center}
\end{figure}
\begin{figure}
\begin{center}
\includegraphics[width=1.0\columnwidth]{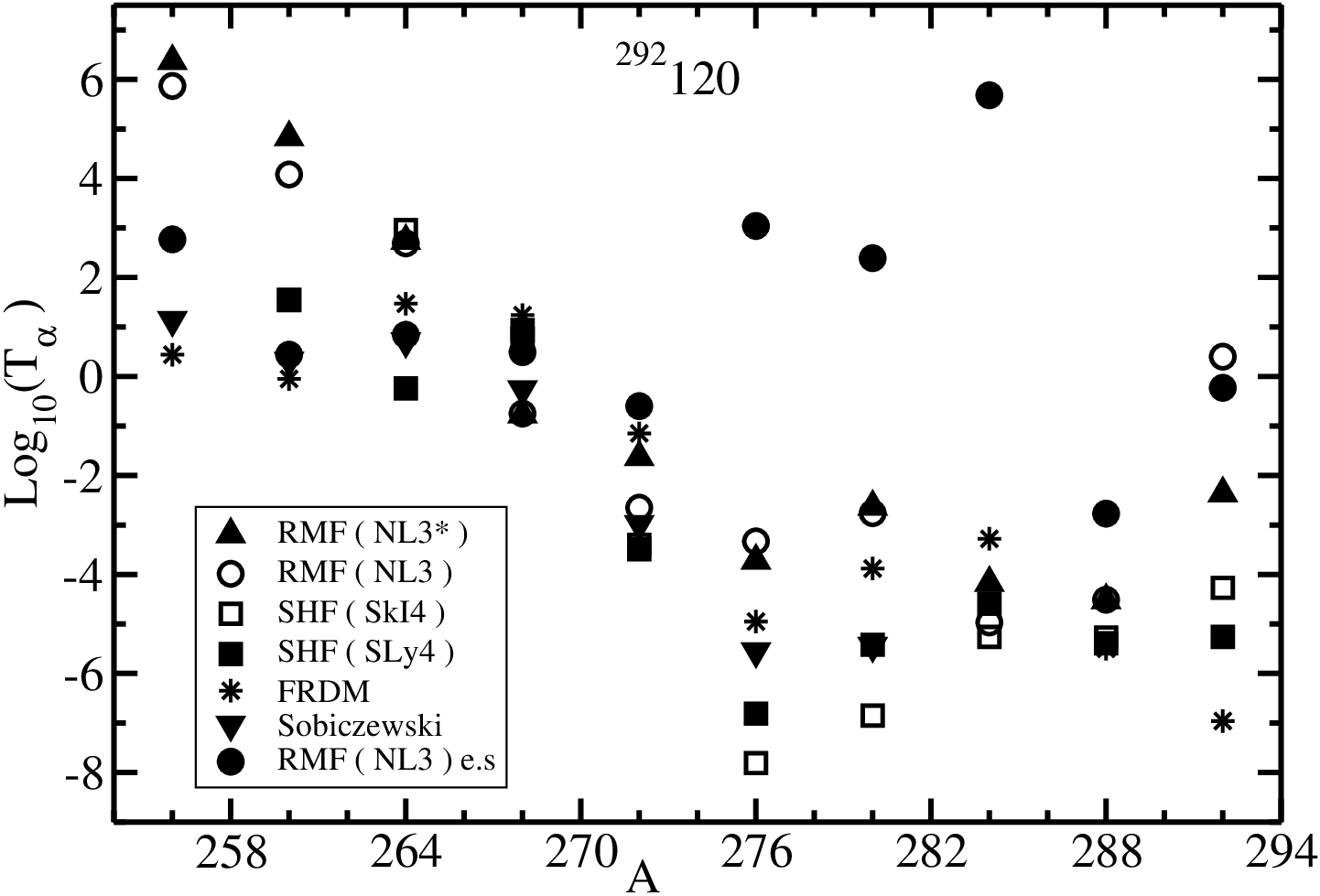}
\caption{The half-life time $T_{1/2}^\alpha$ (in seconds) for the $\alpha$-decay
chain of the $^{292}$120 nucleus using nonrelativistic
SHF(SkI4), SHF(SLy4) and relativistic mean-field formalism
(RMF) with NL3* and NL3 parameter set, compared with the FRDM [62] 
and other theoretical results [29,67,68,69] wherever available.}
\end{center}
\end{figure}

We choose the nucleus $^{292}$120 (N = 172) for illustrating our
calculations of the $\alpha$-decay chain and the half-life time
$T_{1/2}^\alpha$. The binding energies of the parent and daughter nuclei
are obtained by using both the RMF and SHF formalisms.
The $Q_{\alpha}$ values are then calculated; they are shown
in Table 3 and 4 and in Figure 7. Then, the
half-life $T_{1/2}^{\alpha}$ values are estimated by using
the above formulae, and are also given in Table 3 and 4
and in Figure 8. Our predicted results for both $Q_\alpha$ and $T_\alpha$
for the decay chain of $^{292}120$
are compared with the finite range
droplet model (FRDM) calculation \cite{moll97,moll97a}, as well as with
the results of other authors~\cite{patyk91,mu01,muni01,muni03}.

From Figure 7 and 8 and Table 3 and 4, we notice that the
calculated values for both $Q_\alpha$ and $T_\alpha$ agree fairly
well with the FRDM predictions, as well as with the other
theoretical results available. For example, the value of
both $Q_\alpha$ and $T_\alpha$, in both the FRDM and SHF model,
coincides well for the $^{288}118$. For $^{280}114$
isotope, the SHF (SLy4) prediction also coincides well with the
Sobiczewski result for both $Q_\alpha$ and $T_\alpha$, and
the $Q_\alpha$ value of RMF with FRDM result. Furthermore, the
possible shell structure effects in $Q_\alpha$, as well as in
$T_\alpha$, are noticed for the daughter nucleus $^{284}116$ (N = 168)
and $^{292}120$ (N = 172) in RMF (NL3*) and RMF(NL3), $^{262}106$ (N = 156)
and $^{284}116$ (N = 168) in SHF (SkI4), $^{268}108$ (N = 160)
and $^{284}116$ (N = 168) in SHF (SLy4) and $^{284}116$ (N = 168)
in FRDM calculations. Note that these proton and neutron numbers
refer to either observed or predicted magic numbers.

\begin{table}
\caption{The $Q_{\alpha}$ and $log_{10}T_{1/2}^{\alpha}$ (in sec) for
$\alpha$-decay series of $^{304}$120 nucleus, calculated in the SHF
(SkI4), SHF (SLy4), RMF(NL3*) and RMF(NL3) models, and compared with the Finite
Range Droplet Model (FRDM) results [62] and other theoretical
results [29,67,68,69] wherever available.
In order to compare, we have calculated the $T_{1/2}^{\alpha}$
values in case of Sobiczewski {\it et al.}
using Eq.(15).
}
\renewcommand{\tabcolsep}{0.4cm}
\renewcommand{\arraystretch}{0.6}
\begin{tabular}{ccccccccc}
\hline
A & Z & Formalism & BE & $\beta_2$ & $Q_{\alpha}$ & $log_{10}T_{1/2}^{\alpha}$ \\
\hline
304 & 120 & RMF(NL3*) & 2142.57 & 0.59  & 10.01 & 2.14\\
    &     & RMF(NL3)  & 2146.37 & 0.59  & 9.94  & 2.37  \\
    &     &           & 2141.34 & 0.00  & 12.08 & -3.26 \\
    &     & SHF(SKI4) & 2134.96 & 0.00  & 11.95 & -2.95 \\
    &     &           & 2132.61 & 0.56  & 11.93 & -2.95 \\
    &     & SHF(SLy4) & 2112.37 & 0.01  & 11.08 & -0.83 \\
    &     &           & 2108.29 & 0.56  & 11.62 & -2.17 \\
    &     & FRDM      & 2137.99 & 0.00  & 13.82 & -6.83 \\
    &     & Sobiczewski &       & 0.00  & 13.07 & -5.38 \\
300 & 118 & RMF(NL3*) & 2124.29 & 0.58  & 9.29  &3.78  \\
    &     & RMF(NL3)  & 2128.01 & 0.58  & 9.47  & 3.18  \\
    &     &           & 2125.12 & -0.00 & 10.74 & -0.54 \\
    &     & SHF(SKI4) & 2118.60 & 0.00  & 11.02 & -1.27 \\
    &     &           & 2116.24 & 0.56  & 9.82  & 2.06 \\
    &     & SHF(SLy4) & 2095.15 & 0.01  & 9.98  & 1.59 \\
    &     &           & 2091.61 & 0.54  & 10.81 & -0.73 \\
    &     & FRDM      & 2123.51 & 0.00  & 12.72 & -5.18 \\
    &     & Sobiczewski &       & 0.00  & 11.98 & -3.58 \\
296 & 116 & RMF(NL3*) & 2105.28 & 0.54  & 9.33  &2.96  \\
    &     & RMF(NL3)  & 2109.18 & 0.54  & 9.40  & 2.73 \\
    &     &           & 2107.56 & -0.04 & 9.74 & 1.66  \\
    &     & SHF(SKI4) & 2101.32 & 0.03  & 9.88  & 1.25 \\
    &     &           & 2097.76 & 0.55  & 9.36  & 2.85 \\
    &     & SHF(SLy4) & 2076.83 & 0.04  & 9.19  & 3.41 \\
    &     &           & 2074.12 & 0.53  & 9.81  & 1.44 \\
    &     & FRDM      & 2107.94 & -0.01 & 11.10 & -2.08 \\
    &     & Sobiczewski &       & 0.00  & 10.71  & -1.07 \\
292 & 114 & RMF(NL3*) & 2086.31 & 0.51  & 8.77  & 4.14\\
    &     & RMF(NL3)  & 2090.28 & 0.51  & 8.99  & 3.37 \\
    &     &           & 2089.00 & 0.06  & 8.88  & 3.75 \\
    &     & SHF(SKI4) & 2082.90 & 0.03  & 7.62  & 8.59 \\
    &     &           & 2078.82 & 0.51  & 9.07  & 3.12 \\
    &     & SHF(SLy4) & 2057.72 & 0.08  & 9.49  & 1.77 \\
    &     &           & 2055.63 & 0.53  & 9.25  & 2.52 \\
    &     & FRDM      & 2090.75 & -0.02 & 8.25  & 6.02 \\
    &     & Sobiczewski &       & 0.00  & 9.60  & 1.42 \\
288 & 112 & RMF(NL3*) & 2066.78 & 0.49  & 8.83  & 3.22\\
    &     & RMF(NL3)  & 2070.98 & 0.49  & 8.55  & 4.18 \\
    &     &           & 2069.58 & -0.09 & 9.37  & 1.46 \\
    &     & SHF(SKI4) & 2062.22 & 0.10  & 10.04 & -0.51 \\
    &     &           & 2059.59 & 0.52  & 7.12  & 9.99 \\
    &     & SHF(SLy4) & 2038.91 & 0.11  & 8.85  & 3.65 \\
    &     &           & 2036.58 & 0.52  & 8.70  & 3.13 \\
    &     & FRDM      & 2070.70 & -0.06 & 8.34  & 4.92 \\
    &     & Sobiczewski &       & 0.09  & 9.04  & 2.51 \\
\hline
\end{tabular}
\end{table}

\begin{table}
\caption{Same as Table 5.}
\renewcommand{\tabcolsep}{0.40cm}
\renewcommand{\arraystretch}{0.6}
\begin{tabular}{ccccccccc}
\hline
A & Z & Formalism & BE & $\beta_2$ & $Q_{\alpha}$ & $log_{10}T_{1/2}^{\alpha}$ \\
\hline
284 & 110 & RMF(NL3*) & 2047.31 & 0.14  & 7.65  & 6.87\\
    &     & RMF(NL3)  & 2051.23 & 0.13  & 7.88  & 5.92 \\
    &     &           & 2050.65 & 0.48  & 7.34  & 8.11 \\
    &     & SHF(SKI4) & 2043.95 & 0.13  & 8.02  & 5.39 \\
    &     &           & 2038.41 & 0.36  & 11.98 & -5.84 \\
    &     & SHF(SLy4) & 2019.46 & 0.15  & 8.11  & 5.05 \\
    &     &           & 2016.98 & 0.49  & 8.60  & 3.27 \\
    &     & FRDM      & 2050.75 & 0.10  & 7.57  & 7.18 \\
    &     & Sobiczewski &       & 0.10  & 8.34  & 4.19 \\
280 & 108 & RMF(NL3*) & 2026.66 & 0.16  & 7.31  & 7.49\\
    &     & RMF(NL3)  & 2030.81 & 0.15  & 7.72  & 5.77 \\
    &     &           & 2029.69 & 0.44  & 6.73  & 10.14 \\
    &     & SHF(SKI4) & 2023.68 & 0.17  & 7.71  & 5.82 \\
    &     &           & 2022.11 & 0.26  & 8.58  & 2.62 \\
    &     & SHF(SLy4) & 1999.26 & 0.17  & 8.10  & 4.31 \\
    &     &           & 1997.28 & 0.43  & 8.59  & 2.57 \\
    &     & FRDM      & 2030.03 & 0.11  & 7.56  & 6.43 \\
276 & 110 & RMF(NL3*) & 2005.67 & 0.17  & 6.83  & 8.11\\
    &     & RMF(NL3)  & 2010.24 & 0.17  & 6.96  & 8.19 \\
    &     &           & 2008.12 & 0.41  & 8.13  & 3.44 \\
    &     & SHF(SKI4) & 2003.08 & 0.19  & 7.23  & 6.97 \\
    &     &           & 2002.35 & 0.32  & 7.96  & 4.06 \\
    &     & SHF(SLy4) & 1979.07 & 0.20  & 7.64  & 5.31 \\
    &     &           & 1977.58 & 0.40  & 8.74  & 1.34 \\
    &     & FRDM      & 2009.28 & 0.14  & 7.20  & 7.13 \\
\hline
\end{tabular}
\end{table}


\subsection{The $\alpha$-decay series of $^{304}$120 nucleus}
\begin{figure}[t]
\begin{center}
\includegraphics[width=1.0\columnwidth]{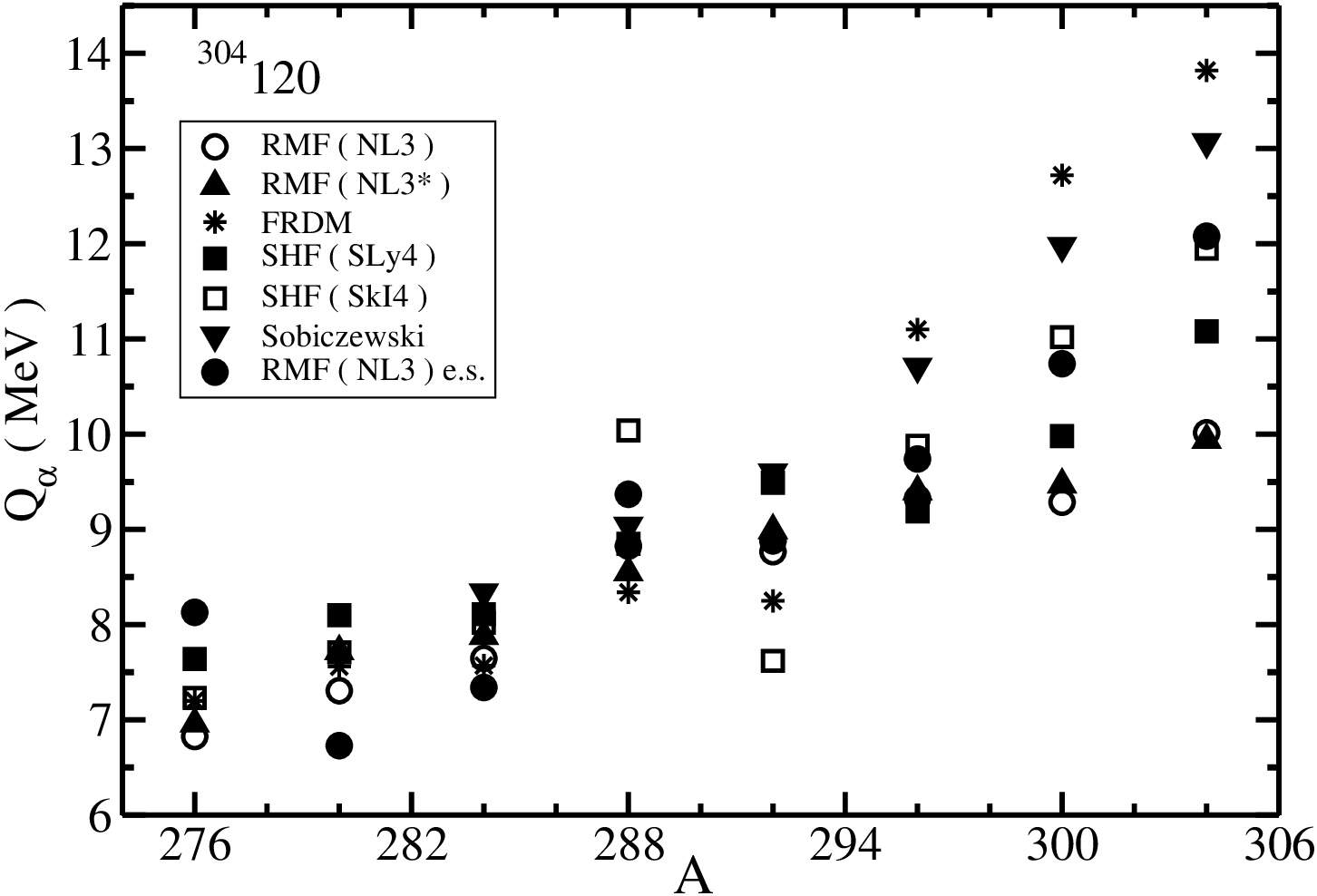}
\caption{The $Q_\alpha$ energy for the $\alpha$-decay
chain of the $^{304}$120 nucleus using nonrelativistic
SHF(SkI4), SHF(SLy4) and relativistic mean-field formalism,
(RMF) with NL3* and NL3 parameter set, compared with the FRDM [62]
and other theoretical results [29,67,68,69] wherever
available.}
\end{center}
\end{figure}

\begin{figure}
\begin{center}
\includegraphics[width=1.0\columnwidth]{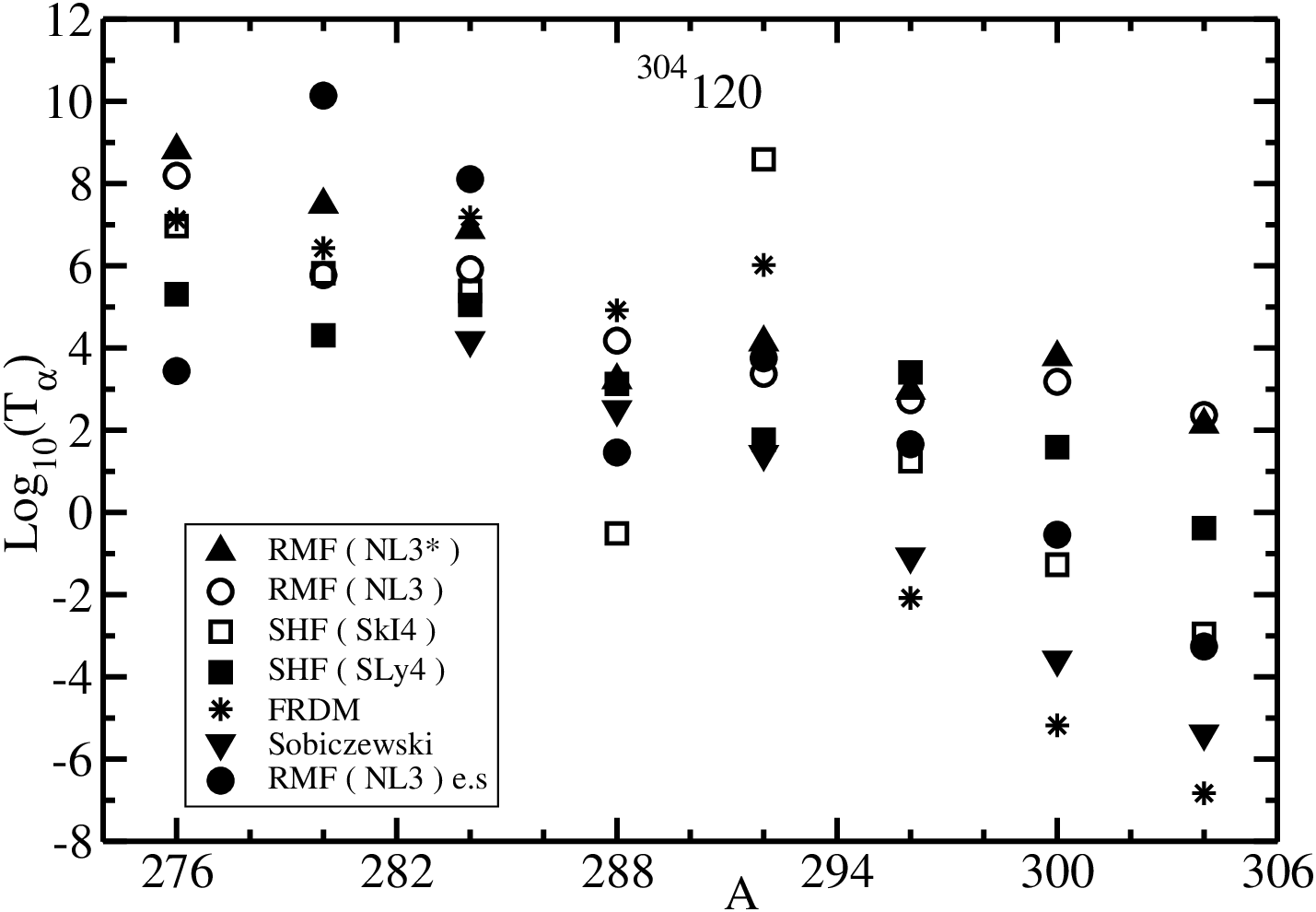}
\caption{The half-life time $T_{1/2}^\alpha$ (in seconds) for the $\alpha$-decay
chain of the $^{304}$120 nucleus using nonrelativistic
SHF(SkI4), SHF(SLy4) and relativistic mean-field formalism
(RMF) with NL3* and NL3 parameter set, compared with the FRDM [62] and other
theoretical results [29,67,68,69] wherever available.}
\end{center}
\end{figure}
In this subsection, we present the $Q_{\alpha}$ and $T_{1/2}^{\alpha}$ results
for the decay series of $^{304}$120 nucleus, using the
same procedure as explained in the previous subsection for
$^{292}$120 nucleus. The results obtained are listed in Table 5 and 6,
and plotted in Figures 9 and 10, compared with the FRDM
predictions~\cite{moll97,moll97a}, as well as with the results
of other authors~\cite{patyk91,mu01,muni01,muni03}.

From Figure 9 and Tables 5 and 6, we notice that the calculated
values for $Q_\alpha$ agree quite well within different models at
different mass numbers. For example, the value of $Q_\alpha$, in the
RMF, SHF and FRDM model, coincides well with the data for
$^{276}$110 (N = 168) and $^{280}$108 (N = 172). Similarly, for
$^{288}$112 (N = 176) and $^{292}$114 (N = 178), the SHF prediction
matches the Sobiczewski {\it et al.} result. For
$^{288}$112 (N = 176) we also have good agreement between RMF and FRDM
result. But towards high mass number we do not have agreement within
different models. From Figures 10 and Table 5 and 6, we can notice almost
similar nature of the half-life time $T_{1/2}^{\alpha}$ values between
different models. Possible shell structure effects in
$Q_\alpha$, as well as in $T_\alpha$, are noticed for the daughter
nucleus $^{284}110$ (N = 174) and
$^{292}114$ (N = 178) in RMF(NL3*),
$^{280}108$ (N = 172) in RMF (NL3), $^{284}110$ (N = 174) and
$^{292}114$ (N = 178) in FRDM and in SHF (SkI4), and $^{280}108$
(N = 172) and $^{284}110$ (N = 174) in SHF (SLy4). Here again
we note that these proton and neutron numbers
refer to either observed or predicted magic numbers.

\section{Summary}
Summarizing, we have calculated the binding energy, and quadrupole 
deformation parameter for the isotopic chain of $^{292}$120 and 
$^{304}$120 superheavy element, which are being planned to synthesize. 
We employed both the SHF and RMF formalisms using various parameter 
sets, for both the ground as well as intrinsic first excited states 
to see the model dependence of the results. We found qualitatively 
similar predictions in both techniques and the results obtained here 
also consistents to earlier calculation with different forces 
\cite{rutz97,rutz99,ren001,ren002,ren03,ren04,zha05}. From the 
calculated binding energy, we also estimated the two-neutron 
separation energy ($S_{2n}$) and the energy difference ($\triangle E$) 
between ground state and first excited state for studying the shape 
coexistence. A shape change from prolate to highly deformed oblate 
at A = 282, and from highly deformed prolate to highly deformed 
oblate at A = 320 is observed in RMF formalism. In RMF calculation 
most of the ground state structures are found to be highly deformed 
prolate, differing strongly with the FRDM calculation where most of 
the ground state structures are with spherical solutions. However, 
in SHF formalism we found that the ground state structures along the 
whole isotopic chain are prolate. Thus, in RMF formalism most of the 
isotopes are superdeformed in their ground state configurations, and 
are low laying highly deformed states in case of SHF formalism. From 
the binding energy analysis, we found that the most stable isotope in 
the $Z$=120 series is around $^{304}$120, which is near to predicted 
magic number at N = 184 \cite{bunu12}. Our predicted $\alpha$-decay 
energy $Q_\alpha$ and half-life time $T_{1/2}^\alpha$ agree nicely 
with the FRDM and other available theoretical results 
\cite{patyk91,mu01,muni01,muni03}. 

\section*{Acknowledgments}
S. Ahmad and M.Bhuyan would like to thank Institute of Physics for local
hospitality during the course of this work. This work is supported in part
by the Council of Scientific $\&$ Industrial Research, HRDG, CSIR Complex,
Pusa, New Delhi-110012, India (Project No. 09/153 (0070)/2012 EMR-I.

\end{document}